%% file: ms.tex
\documentclass[runningheads]{llncs}
\usepackage{cite}
\usepackage{amsmath,amssymb,amsfonts}
\usepackage{graphicx}
\usepackage{multirow}
\usepackage{hyperref}
\usepackage{wrapfig}
\usepackage{endnotes}
\let\footnote=\endnote
\usepackage{color}    
\usepackage[colorinlistoftodos,prependcaption,textsize=scriptsize]{todonotes} 
\usepackage{xargs}  

\newcommandx{\phh}[2][1=]{\todo[linecolor=blue,backgroundcolor=white,bordercolor=blue,#1]{#2}}
\newcommandx{\rs}[2][1=]{\todo[linecolor=red,backgroundcolor=white,bordercolor=red,#1]{#2}}

\begin{document}

\expandafter\def\expandafter\UrlBreaks\expandafter{\UrlBreaks
  \do\a\do\b\do\c\do\d\do\e\do\f\do\g\do\h\do\i\do\j%
  \do\k\do\l\do\m\do\n\do\o\do\p\do\q\do\r\do\s\do\t%
  \do\u\do\v\do\w\do\x\do\y\do\z\do\A\do\B\do\C\do\D%
  \do\E\do\F\do\G\do\H\do\I\do\J\do\K\do\L\do\M\do\N%
  \do\O\do\P\do\Q\do\R\do\S\do\T\do\U\do\V\do\W\do\X%
  \do\Y\do\Z\do\0\do\1\do\2\do\3\do\4\do\5\do\6\do\7%
  \do\8\do\9}

\title{Mutation testing of smart contracts at scale}

\author{Pieter Hartel\inst{1,2} \and Richard Schumi\inst{3}}

\authorrunning{Hartel and Schumi}

\institute{Singapore University of Technology and Design \and
	Delft University of Technology \and
	Singapore Management University}

\maketitle

\begin{abstract}

It is crucial that smart contracts are tested thoroughly due to their immutable 
nature. Even small bugs in smart contracts can lead to huge monetary losses. 
However, testing is not enough; it is also important to ensure the quality and 
completeness of the tests. There are already several approaches that tackle 
this challenge with mutation testing, but their effectiveness is questionable 
since they only considered small contract samples. Hence, we evaluate the 
quality of smart contract mutation testing at scale. We choose the most 
promising of the existing (smart contract specific) mutation operators, analyse 
their effectiveness in terms of killability and highlight severe 
vulnerabilities that can be injected with the mutations. Moreover, we improve 
the existing mutation methods by introducing a novel killing condition that is 
able to detect a deviation in the gas consumption, i.e., in the monetary value 
that is required to perform transactions.
\keywords{Mutation testing, Ethereum, Smart contracts, Solidity, Gas limit as a killing criterion, Vulnerability injection, Modifier issues.}
\\
{\bf This paper has a replication package}\footnote{Replication package of this paper \url{https://github.com/pieterhartel/Mutation-at-scale}}
\end{abstract}

\section{Introduction}
\label{sec:introduction}
\input{sec/introduction}

\section{Background}
\label{sec:background}
\input{sec/background}

\section{Method}
\label{sec:method}
\input{sec/method}

\section{Results}
\label{sec:results}
\input{sec/results}

\section{Discussion and Limitations}
\label{sec:discussion}
\input{sec/discussion}

\subsection{Limitations and Threats to Validity}
\label{sec:limitations}
\input{sec/limitations}

\section{Conclusions and future work}
\label{sec:conclusions}
\input{sec/conclusions}


\renewcommand{\notesname}{Web references}
\theendnotes

\bibliographystyle{plainurl}
\bibliography{darkweb_refs,richard_refs}

\end{document}

%% file: sec/introduction.tex
Smart contracts are programs designed to express business logic for managing the data or assets on a blockchain system.
Although smart contracts already exist for some years, they still suffer from security vulnerabilities, which can lead to huge monetary losses~\cite{Atzei2017}.
Hence, it is crucial to make sure that smart contracts do not contain such vulnerabilities.
The most important method for finding both vulnerabilities and semantic errors is testing.
Testing smart contracts is even more essential than testing regular programs, since their source is often publicly available, which makes them an easy target, and updating them is cumbersome due to their immutable nature.
Moreover, it is critical to ensure the quality of the tests.
There are a few quality metrics, like code coverage, i.e., the percentage of the source code that is executed by a test, but code coverage is not able to measure the error detection capability of tests and it is rarely a good indicator for the number of faults in a software\cite{DBLP:conf/icst/TengeriVBJBVG16}. 
A technique that can perform such measurements is mutation testing, which injects faults into a program to check if the tests can detect these faults.

There have already been a number of publication that showed mutation approaches for Solidity\footnote{Solidity documentation \url{https://solidity.readthedocs.io}} smart contracts
\cite{Andesta2019, Bugrara2019, Chapman2019, Fu2019, Groce2018, Peng2019, Wang2019a, Wang2019b, Honig2019, Wu2019a}.
Solidity is a JavaScript like language\footnote{On the differences between Solidity and Javascript \url{https://vomtom.at/whats-the-difference-between-javascript-solidity-and-ethereum}} 
with several special features to interact with the underlying Ethereum blockchain.
The blockchain stores code and data, and it is managed by the owners of the Ethereum peer-to-peer network.
Many of the related mutation testing approaches introduced interesting smart contract specific mutations, but they only performed small evaluations with a few contracts.
We selected the most promising mutation operators of the related work, generalized them, and performed a large scale evaluation with about a thousand contracts for a meaningful quality assessment of the operators.
We use replay tests downloaded from Truffle-tests-for-free~\cite{Hartel2019a} that are automatically produced from historic transaction data on the blockchain.
The achieved mutation score can serve as a baseline for testing other, more sophisticated testing methods.


We are particularly interested in smart contract specific mutations that simulate common mistakes made by smart contract developers.
An example is a forgotten or wrong function modifier~\cite{Honig2019}.
A modifier can express conditions that have to be fulfilled for the execution of a function, e.g., that the caller of the function is the owner of the contract.
Since modifiers are often concerned with access control, omitting a modifier can have catastrophic effects.
For example management functions of a smart contract can become publicly available.

Another smart contract specific aspect is the gas consumption of transactions.
Everything on Ethereum costs some units of gas~\cite{Wood2017}.
For example, executing an \verb=ADD= bytecode costs 3 gas.
Storing a byte costs 4 or 68 gas, depending on the value of the byte (zero or non-zero).
The price of gas in Ether varies widely\footnote{Gas price tracking \url{https://etherscan.io/chart/gasprice}}, and the market determines the exchange rate of Ether.
The cost of a transaction can be anything from less than a cent to several US\$.
Executing smart contracts is therefore not just a matter of executing the code with the right semantics but also of cost control.
Therefore, all transactions have a gas limit to make sure that the cost is managed.
Executing smart contracts with a gas limit is comparable to executing code on a real time system with a deadline~\cite{Nilsson2006}.
This opens up new possibilities for killing mutants, over and above the standard killing conditions.
Similar to detecting mutants on real-time systems with a different timing behaviour, we measure the gas consumption of tested transactions to find deviations to reference executions of these transactions.
This allows us to kill mutants that consume a significantly greater amount of gas. 

Our major contributions are:
(1) We propose a set of mutation operators on the basis of related work and evaluate these operators at scale.
(2) To further improve the mutation score, we introduce a novel killing condition based gas limits for smart contract transactions.

%% file: sec/background.tex
Mutation testing~\cite{Offutt2001,Jia2011,Papadakis2019} is an evaluation technique for assessing the quality of a set of test cases (i.e., a test suite).
It works by introducing faults into a system via source code mutation and by analysing the ability of the test suite to detect these faults.
The idea is that the mutation should simulate common mistakes by developers.
Hence, when a test suite is able to find such artificial faults, it should also find real faults that can occur through programming mistakes.

Developers are likely to make mistakes with standard language features, but because Ethereum is relatively young, they are more likely to confuse Solidity specific features.
For example, Solidity offers two different types of assertions: \verb=require(.)= is used to check external consistency, and \verb=assert(.)= is used to check internal consistency.
Both terminate the contract but with a different status.

Developers also have trouble with the qualifiers that Solidity offers, for example \verb=external= is for functions that can be called from other contracts and via transactions, but not internally, and \verb=public= is for functions that can either be called internally or via transactions.

Finally, the addresses of contracts and externally owned accounts play such an important role in smart contracts that there are several ways of specifying addresses that may confuse the developer.
For example \verb=msg.sender= is the address of the sender of a message, and \verb=tx.origin= is the address of the externally owned account that sent a transaction.
They are the same for a short call chain but not for a longer call chain.



Mutation testing is an old technique, but it has still open challenges, like the equivalent mutant problem, which occurs when a mutation does not change the original program, e.g., when a fault is injected in dead code.
There are methods to detect equivalent mutants~\cite{DBLP:journals/stvr/OffuttC94,DBLP:conf/icst/GrunSZ09,DBLP:journals/stvr/HieronsHD99,DBLP:journals/stvr/OffuttP97}, but it is still not possible to remove all equivalent mutants.
Hence, this limits the usability of mutation testing, since a high manual effort is required to identify equivalent mutants.
There are 11 related papers that propose mutation testing operators for Solidity.
The number of introduced mutation operators in these publications varies widely, since it is up to the tester to choose the scope or specificity of the operators. 
Some authors prefer to introduce a specific operator for every singular change, others choose to group together similar changes into one operator, which is more common and was also done by us.

Bond\footnote{There is no paper available on eth-mutants, but there is a GitHub page \url{https://github.com/federicobond/eth-mutants}} implements just one mutation operator from the Mothra set and does not provide an evaluation.
Burgrara~\cite{Bugrara2019} does not mutate Solidity, but manually mutates lower level EVM code, ABI encodings and public key operations.
Chapman~\cite{Chapman2019} proposes 61
mutation operators for Solidity and evaluates them on a set of six DApps.
Fu et al.~\cite{Fu2019} propose mutation testing for the implementation of the Ethereum Virtual Machine (EVM), but not for smart contracts.
Groce et al.~\cite{Groce2018} describe a generic mutation tool with a set of specific operators for Solidity, but without an evaluation.
Peng et al.~\cite{Peng2019} describe five mutation operators and evaluate them on a set of 51 smart contracts.
Wang et al.~\cite{Wang2019a} use some unspecified mutations from the Mothra set to study test coverage.
Wang et al.~\cite{Wang2019b} do not mutate Solidity but transactions sequences.

Three papers are closely related to ours and served us as a basis for our mutation operators:
Andesta et al.~\cite{Andesta2019} propose 57 mutation operators for Solidity and evaluate them by investigating how the mutation operators are able to recreate known attacks, such as the DAO attack~\cite{Mehar2019}. 
The authors do not provide mutation scores, and they only evaluate to what extent they can reproduce known vulnerabilities in a few contracts. Hence, they show no evaluation for most of their operators.
Honig et al.~\cite{Honig2019} describe two Solidity specific operators and adopt four existing operators. 
They evaluate the operators on two popular DApps that have extensive test suites with high code coverage. These test suites allow them to achieve high mutation scores, but the scope of their mutations is limited.
Wu et al.~\cite{Wu2019a} propose 15 Solidity specific operators, which were also supported by their tool called MuSC\footnote{There was a tool demo at ASE 2019 without a paper, but there is a GitHub page \url{https://github.com/belikout/MuSC-Tool-Demo-repo}},
and tested the operators on four DApps. 
They evaluate their mutation approach by comparing the effectiveness of a test suite that was optimised based on the mutation score to one that was optimised based on code coverage. Moreover, they also point out vulnerabilities that can be simulated with their operators. In contrast to our work, they have fewer operators concerning access control and hence they cannot reproduce some severe vulnerabilities regarding unauthorized access.

With our generalised mutation operators we are able to inject nearly all the changes from related work, with a few minor exceptions. For example, we do not mutate data types because it causes too many compilation errors.
The evaluation of related work is limited to just a few DApps, and the results vary.
The research question that follows from the analysis above is:
{\it How efficient are the standard mutation operators as compared to Solidity specific operators?}

To break this question down into its more manageable sub questions we present a case study in mutation testing of a sample smart contract first, and then list the sub questions.

\subsection{A case study in mutation: Vitaluck}
\label{subsec:case_study}
As a case study we use a lottery contract called Vitaluck~\cite{Chia2018}.
The source of the contract can be browsed on Etherscan\footnote{Vitaluck on Etherscan \url{https://etherscan.io/address/0xef7c7254c290df3d167182356255cdfd8d3b400b}}.
The contract contains a main method called \verb=Play= and a number of management methods; \verb=Play= contains the core of the business logic of the lottery.
Each call to \verb=Play= draws a random number in the range 1 to 1000 using the time stamp of the current block as a source of entropy.
If the random number is greater than 900, the player wins the jackpot, and a percentage of each bet is paid to the owner of the contract.

Vitaluck is a relatively short contract (139 lines of source code excluding comments).
It has not been used extensively; there are only 27 historic transactions that can also be browsed on Etherscan.
The first transaction deploys the contract, and the remaining historic transactions are all calls to the \verb=Play= method.
None of management methods of the contract are ever called by the historic transactions on the blockchain.
However, the \verb=Play= method occupies the majority of the code and provides ample opportunities for using standard and Solidity specific operators.
We give a number of examples of mutations below.

Each example is labelled with the mutation operator and a brief description of the operator.
We indicate common known vulnerabilities, as described in the smart contract weakness classification (SWC) registry~\cite{swc}, which can be simulated with the mutation operators.
\autoref{tab:operator_defs} summarises the operators.

{\it LR\_I - Literal Integer replacement}
Since Vitaluck is a lottery, any mutation to the code that manages the jackpot has a high likelihood of causing a fault in the contract.
The first sample mutation (line 149) changes \verb=900= to \verb=1=.
This is shown below, using an output format inspired by the Unix \verb=diff= command.
The range of \verb=_finalRandomNumber= is 1 to 1000.
If the condition in the \verb=if= statement is true, the jackpot will be paid out, which in the original code happens on average 10\% of the time.
After the mutation, the jackpot will be paid out 99.9\% of the time, which completely breaks the contract.
{\footnotesize%
\vspace{-1ex}
\begin{verbatim}
< if(_finalRandomNumber >= 900) {
> if(_finalRandomNumber >=  1 ) {
\end{verbatim}
\vspace{-1ex}
}

To determine if a replay test kills a mutant, we compare the output of the original contract to the output of the mutant.
The output of a contract consists of the status and the emitted events of the transactions, and the outputs of the pure functions called by the test.
The mutant above does not affect the status of any of the transactions of the test (they all succeed), but it does cause the event \verb=NewPlay(address player, uint number, bool won);= to be emitted 17 times more with \verb!won=true! than the original contract.
Similarly, the mutant causes the pure function \verb=GetWinningAddress()= to return a different address than the original contract.
Both these differences are easy to detect from the output of the replay test, thus supporting the conclusion that this particular mutant is killed.
For the remaining examples, we will not discuss the outputs.



{\it MORD* -  Modifier Replacement or Deletion} The sample mutation below (line 257) deletes the modifier \verb=onlyCeo= from the method that installs the address of a new CEO.
This allows anyone to set the payout address to his own, rather than just the CEO.
This behaviour also corresponds to the vulnerability SWC-105, which can occur when the access control for functions is insufficient.
The operator can further cause vulnerabilities, like the SWC-106 Unprotected SELFDESTRUCT Instruction, or SWC-123 Requirement Violation. Note that adequate tests for such faults would try to call these functions with unauthorized users in order to check if the expected error message occurs. 
{\footnotesize%
\vspace{-1ex}
\begin{verbatim}
< function modifyCeo(address _newCeo) public onlyCeo {
> function modifyCeo(address _newCeo) public {
\end{verbatim}
\vspace{-1ex}
}

{\it BOR - Boolean Operator Replacement }
The last sample mutation (line 93) replaces the Boolean operator \verb!=! in the second statement of the function \verb=Play=.
{\footnotesize%
\vspace{-1ex}
\begin{verbatim}
< if(totalTickets  = 0) {  ... return; }
> if(totalTickets != 0) {  ... return; }
\end{verbatim}
\vspace{-1ex}
}

Most smart contracts use the constructor to initialise the state of the contract.
For some unknown reason, \verb=Vitaluck= does not have a constructor.
Instead, the contract relies on the first call to \verb=Play= to initialise the state, including the jackpot.
This is poor coding style, and it may be a security problem too.
The mutant above allows us to discover the problem as follows.
The first \verb=Play= transaction executes the then branch, for which it needs 62347 gas out of a gas limit of 93520.
However, the mutant skips the then branch and executes the rest of the \verb=Play= method.
This takes 272097 gas, which is about 3 times the gas limit.
This suggests that using the gas limit to kill mutants might be of interest.



\subsection{Sub questions} 	
Based on the Vitaluck case study we formulate subsidiary research questions.

{\bf Discarding stillborn mutants}
We created a simple tool (ContractMut) that makes maximum use of existing state-of-the-art tools, such as the Truffle framework\footnote{Truffle framework documentation \url{https://www.trufflesuite.com}} and the Solidity compiler.
In particular, the tool relies heavily on the Solidity compiler to read the source file and to generate the abstract syntax tree (AST).
It would also be possible to mutate at the bytecode-level, but this would make it more difficult to understand what a mutant is doing. 
The AST approach has the advantages that the amount of bespoke tooling to be built is limited.
The disadvantage is that the compiler has more information than it exposes via the AST.
Our mutation tool does not have semantic information about the original code or the mutant as it works on the AST generated by the Solidity compiler.
This means that some mutants are generated that do not compile.
For example, when replacing \verb=*= by \verb=-= in an expression such as \verb=1 * 9 / 10000000000000000000=, an error occurs because a rational constant cannot be subtracted from an integer.
This raises the sub question: {\it To what extent does the tool generate stillborn mutants?}

{\bf Discarding duplicate and equivalent mutants}
Duplicate and equivalent mutants distort the mutation score.
The next sub question is therefore: {\it How to detect duplicate and equivalent mutants?}

{\bf Mutation score}
The purpose of generating mutants is to assess the quality of a test.
If too few mutants are killed by a test, the usual course of action is to develop more tests, which is labour-intensive.
However, for a smart contract like Vitaluck that has already been deployed on the blockchain, all historic transactions are available and can be decompiled into a test.
A test then consists of replaying the sequence of historic transactions.
We can control the size of the test by varying the number of transactions executed by the mutant.
As the size of the test increases, we expect the mutation score to increase.
Unfortunately, most historic transactions execute the same method calls, so the increase in mutation score should tail off rapidly.
The sub question then is: {\it What is the relation between the mutation score and the length of the test?}

{\bf Efficiency of the mutation operators}
We follow the competent programmer hypothesis~\cite{Zhu2018} by mutating a subtle variant of a program fragment that could have been created by mistake.
The sub question is: {\it What is the relative success of standard mutation operators as compared Solidity specific operators?}


{\bf Using outputs in a mutant killing condition}
To determine whether a mutant is killed, a test compares the output of the original to the output of the mutant and kills the mutant if there are difference.
The next sub question is: {\it Which observable outputs can be used in the killing condition?}

{\bf Using gas in a mutant killing condition}
A tight limit on the amount of gas reduces the risk of having to pay too much for a transaction.
However, if the limit is too tight, some method calls may fail unexpectedly.
Deciding on the gas limit is therefore a non-trivial problem.
This raises the following sub question: {\it To what extent is killing mutants based on exceeding the gas limit efficient?}

%% file: sec/method.tex
\begin{center}
\begin{table*}
\caption{Mutation operators for Solidity programs. Operators marked with an asterisk are Solidity specific.}
\begin{tabular*}{1\textwidth}{ l l l l }
\hline
\hline
Mothra		& Our		& Description					& SWC ID	\\ 
operator	& operator	&						&		\\ 
\hline
AOR/LCE/ROR	& AOR		& Assignment Operator Replacement		& 129		\\ 
AOR/LCR/ROR	& BOR		& Binary Operator Replacement			& 129		\\ 
SDL		& ESD		& Expression Statement Deletion			&		\\ 
SVR		& ITSCR		& Identifier with same Type, Scope,		& 105,106	\\ 
		&		& and Constancy Replacement			&		\\ 
RSR		& JSRD		& \multicolumn{2}{l}{Jump Statement Replacement/Deletion}	\\ 
-		& LR\_A*	& Literal Address Replacement			& 115		\\ 
CRP/CSR/SCR/SRC	& LR\_\{B,I,S\}	& Boolean, Int, String Replacement		&		\\ 
-		& MORD*		& Modifier Replacement/Deletion			& 105,106,123	\\ 
-		& QRD*		& Qualifier for storage local or state		& 100,108	\\ 
		&		& mutability Replacement/Deletion		&		\\ 
-		& RAR*		& R-Value Address Replacement			& 115		\\ 
AOR/LCR/ROR	& UORD		& Unary Operator Replacement/Deletion		& 129		\\ 
SVR		& VDTSCS	& Variable Declaration with same Type		&		\\ 
		&		& Scope and Constancy Swap			&		\\ 
\hline
\hline
\end{tabular*}
\label{tab:operator_defs}
\end{table*}
\end{center}
We describe an experiment in mutation testing of smart contracts on Ethereum.
The experiments have been conducted on a uniform random sample of 1,120 smart contracts with tests from Truffle-tests-for-free that can also be downloaded from the replication package of this paper\endnotemark[1].
We removed 157 contracts from the set because either they were non-deterministic or they did not have a test with 50 transactions.
These 963 contracts are representative for the entire collection of 50,000+ verified smart contracts on Etherscan~\cite{Hartel2019a}, and the sample is relatively large~\cite{Papadakis2019}.
The tests are replay tests and relatively short, with an average bytecode coverage of 51.4\%~\cite{Hartel2019a}.
Our results are therefore a baseline.

A test for a contract begins by compiling and re-deploying the contract on a pristine blockchain.
We use the Truffle framework for this with the exact same time stamps and transaction parameters as the historic deployment on the public Ethereum blockchain.
All mainnet addresses are replaced by testnet addresses and all externally owned accounts have a generous balance.
After re-deployment, the first 50 historic transactions are re-played, also with the historic time stamps and transaction parameters~\cite{Hartel2019a}.
After each transaction all pure methods of a contract are called, with fuzzed parameters.
This is intended to simulate any actions by a Distributed Application (DApp) built on top of the contract.

To detect whether a mutant is killed, we compare the outputs of a transaction generated by the original contract to the outputs of the corresponding transaction generated by the mutant contract.
We compare only observable outputs of a transaction, which means that we consider only strong mutants that propagate faults to the outputs.

{\bf Discarding stillborn mutants}
ContractMut uses the Solidity compiler to compile all the mutants it generates.
If the compilation fails, the mutant is discarded.
Smart contracts are usually relatively small, hence the time wasted on failed compilations is limited.

{\bf Discarding duplicate and equivalent mutants}
ContractMut implements the trivially equivalent mutant detection method~\cite{Kintis2018} to discarded duplicate and equivalent mutants.
Each new mutant is compiled and the bytecode of the new mutant is compared to the bytecode of the original and the bytecode of all previously generated mutants.
If there is a match, the new mutant is discarded.

{\bf Mutation score}
Since the tests are machine generated, they are consistent in the sense that all sample contracts are tested by making 50 transactions.
The advantage of using machine-generated tests is that this scales well to large numbers of contracts.
The disadvantage is that the tests are not necessarily representative for handcrafted tests.
For example, the bytecode coverage of the tests varies considerably, from 6\% to 98\%~\cite{Hartel2019a}.

{\bf Efficiency of the mutation operators}
ContractMut implements the core of the Mothra set~\cite{King1991}, which is considered the minimum standard for mutation testing~\cite{Papadakis2019}, as well as the essence of recently proposed Solidity specific operators.
\autoref{tab:operator_defs} lists the operators in alphabetical order.
The first column of the table indicates the correspondence with the Mothra operators, and operators marked with an asterisk are Solidity specific.
The next two columns give the name of the operator, and a description.
The last column shows the relation of the operators to known vulnerabilities based on the smart contract weakness classification (SWC) registry~\cite{swc}.

We have implemented 84.8\% of all 191 mutation operators from related work as a manageable set of 14 operators.
We have not implemented object oriented operators (5.2\%), and leave this for future work.
Signed integers are rare in contracts; hence we have not implemented related operators (1.6\%).
We did not insert mutations at random locations in the code (6.8\%), or type level mutations (1.0\%), because these would generate mostly stillborn mutants.
Our BOR operator covers 30.4\% of the operators from related work, followed by QRD (18.3\%), and ITSCR (11.0\%).
The replication package of this paper\endnotemark[1] provides a table (\verb=comparison.xslx=) mapping the operators from related work onto ours.

Mutants are created as follows.
Each node in the AST represents a program fragment that could in principle be mutated.
Therefore, all relevant AST nodes are collected in a candidate list.
This includes simple statements, literals, identifiers, function parameters, and operators.
Compound statements, methods, and even entire contracts are not in the candidate list, because mutations to such large program fragments are not consistent with the competent programmer hypothesis.
Once the candidate list has been built, the tool repeatedly selects a mutation candidate uniformly at random from the list~\cite{Zhang2010a} and applies the appropriate mutation operator from \autoref{tab:operator_defs}.


\begin{center}
\begin{table}
\caption{Gas used by the historic transaction as a percentage of the historic gas limit versus gas used by the replayed transaction as a percentage of the calculated gas limit.}
\centering
\begin{tabular*}{0.685\textwidth}{l *{4}{r} }
\hline
\hline
			& \multicolumn{2}{c}{Historic}
							& \multicolumn{2}{c}{Replay} \\
\% gas used of limit	& (count)	& (\%)		& (count)	& (\%) \\
\hline
$\geq 0\%  ~\& <20\%$	& 5062		& 11.3\%	& 4292		& 12.3\% \\
$\geq 20\% ~\& <40\%$	& 6893		& 15.4\%	& 5439		& 15.6\% \\
$\geq 40\% ~\& <60\%$	& 6405		& 14.3\%	& 4314		& 12.4\% \\
$\geq 60\% ~\& <80\%$	& 9847		& 21.9\%	& 7522		& 21.6\% \\
$\geq 80\% ~\& <100\%$	& 9564		& 21.3\%	& 9119		& 26.2\% \\
$=100\%$		& 7127		& 15.9\%	& 4171		& 12.0\% \\
Tx success		& 44898		& 100.0\%	& 34857		& 100.0\% \\
Tx failed		& 6032		&		& 16073		& \\
\hline
Total Tx		& 50930		&		& 50930		& \\
\hline
Minimum gas limit	& 21000		&		& 21000		& \\
Maximum gas limit	& 8000029	&		& 15279099	& \\
Mean gas limit		& 397212	&		& 416350	& \\
Std. deviation gas limit& 1002691	&		& 1071615	& \\
\hline
\hline
\end{tabular*}
\label{tab:gas_used}
\end{table}
\end{center}

{\bf Using outputs in a mutant killing condition}
To determine whether a mutant is killed the outputs of the original contract are compared to the outputs of the mutant while executing each transaction of the test.
{\it TxEvMeth} indicates a comparison of all observable outputs of a transaction as follows.
{\it Tx} compares the transaction status (i.e., success, failure, or out of gas).
{\it Ev} compares all outputs of all events emitted by a transaction.
{\it Meth} compares all outputs of all pure methods called by the DApp simulation after each transaction.
The combination {\it TxEvMeth} is the standard mutant killing condition.

{\bf Using gas in a mutant killing condition}
To assess how tight the gas limits on historic transactions are, we have analysed the statistics of all $N=50,930$ historic transactions downloaded from Truffle-tests-for-free.
Columns two and three of \autoref{tab:gas_used} show that 15.9\% of the sample use exactly the gas limit, and 11.3\% use less than 20\% of the gas limit.
The minimum gas limit is 21,000 and the maximum is 8,000,029, which represents a large range.
The standard deviation is also relatively large (1,002,691).
Hence, there is considerable variance in how developers estimate the limit on transactions.

There are two reasons why the gas limit is often loose.
Firstly, the standard tool~\verb=estimateGas= (from \verb=web3.eth=) has to work out which EVM instructions a transaction will execute.
Since each instruction costs a known amount of gas~\cite{Wood2017}, the total gas cost of the transaction can then be calculated.
However, for any non-trivial transaction the exact list of EVM instructions depends on the data in storage, and the data passed as parameters etc.
This makes the estimates unreliable.
\footnote{What are the limitations to estimateGas and when would its estimate be considerably wrong? \url{https://ethereum.stackexchange.com/questions/266}}).

Secondly, since gas costs real money, ultimately the developer has to decide on the basis of the gas estimate what the gas limit of the transaction should be.
For example, setting the gas limit lower than the estimate reduces the risk of losing money via gas-based attacks, but also increases the risk of failing transactions.

For every transaction of a mutant, we could have called~\verb=estimateGas= to obtain an up-to-date gas estimate.
However, we cannot go back to the developer and ask him to decide whether to increase or decrease the gas limit.
In general, we do not even know who the developer might be.
Therefore, we have developed a heuristic that transfers the developers decision on the gas limit of the historic transaction to the gas limit of the mutant transaction.

Assume that the limit as provided by a historic transaction, $\mathit{glh}$, is a hard limit on the amount of gas that the developer is prepared to use.
Then, in principle we can use this limit to kill all mutants executing the same transaction that exceed the limit.
However, since we are replaying each historic transaction on the Truffle framework, the amount of gas used by replaying the transaction may be slightly different.
To compensate for this, we propose to calculate the gas limit on replaying a transaction, $\mathit{glr}$, as the maximum of the gas limit of the historic transaction, and the scaled gas limit of the historic transaction.
The scaling applied is the ratio of $\mathit{gur}$, the gas used by the replay, and $\mathit{guh}$, the gas used by the historic transaction:
\[
\mathit{glr} = \mathit{max}( \mathit{glh}, \frac{\mathit{gur}}{\mathit{guh}} \mathit{glh} )
\]
Here $\mathit{glh}$, and $\mathit{guh}$ are both obtained from historic transaction on the blockchain.
$\mathit{gur}$ is obtained by replaying the historic (i.e. not mutated) transaction on the Truffle framework, with the maximum gas limit.

With this heuristic the last sample mutant of \autoref{sec:background} will be killed.
If we had used \verb=web3.eth.estimateGas= instead of the heuristic, the mutation would not have been killed.
However, this would go against the intention of the developer, who had anticipated that the first call of the \verb=Play= method should just initialise the contract, thus never taking a large amount of gas.

The last two columns of \autoref{tab:gas_used} show the statistics of $\mathit{glr}$ calculated according to the formula above.
The distribution is similar to that of the gas used by the historic transaction, but there is more variance.
The $\mathit{max}$ operation in particular makes the limit on the gas used by the replay less tight than the gas limit on the historic transaction.
In the next section, we will investigate to what extent $\mathit{glr}$ is efficient as a killing condition.
We will call this the {\it Limit} condition, and apply it on its own, and in combination with the other two conditions.

%% file: sec/results.tex
This section describes the results of our experiment in mutation testing of smart contracts on Ethereum.
For each smart contract with a test we tried to generate exactly 50 non-equivalent mutants.
Since each attempt requires a call to the Solidity compiler, we set an upper limit of 1,000 on the number of attempts to generate a mutant.
For 18 smart contracts fewer than 50 trivially non-equivalent mutants were generated, but for the remaining 98.3\% of the contracts we obtained 50 non-equivalent mutants.
In total, we generated 71,314 mutants, of which 47,870 were compilable and not trivial equivalent~\cite{Kintis2018} or duplicate.
For each contract we then executed the tests against all trivially non-equivalent mutants on the Truffle framework.
We ran 47,870 $\times$ 50 = 2,393,500 transactions, which took over a week to run on 14 Linux virtual machines (Xeon dual core, 2.4 GHZ, with 16GB RAM).

{\bf Discarding stillborn mutants}
Of the generated 71,314 mutants, 11,252 (15.8\%) could not be compiled.
As expected the {\it QRD*} operator generates the most stillborn mutants: 63.3\% of the mutants with this operator did not compile.
We take this as an indication that using the limited amount of semantic information available in the Solidity AST is an acceptable approach towards building a baseline mutation tool.
The percentage of stillborn mutants can be reduced to zero if the full power of various semantics analyses of the compiler could be leveraged, but the cost of building such a mutation tool just to reduce a relatively small percentage of failed compilations alone would not be justifiable.

{\bf Discarding duplicate and equivalent mutants}
Of the 71,314 generated mutants, 12,192 (17.1\%) were trivially equivalent to the original or a duplicate of another mutant.
The trivial equivalent detection method~\cite{Kintis2018} that we used is therefore reasonably effective, especially since often about 40--45\% of the mutants can be equivalent~\cite{DBLP:conf/icst/GrunSZ09,DBLP:conf/icst/SchulerZ10}.

\begin{figure}[t]
  \centering
    \includegraphics[width=0.7\textwidth, trim={3mm 0mm 0mm 0mm}, clip]{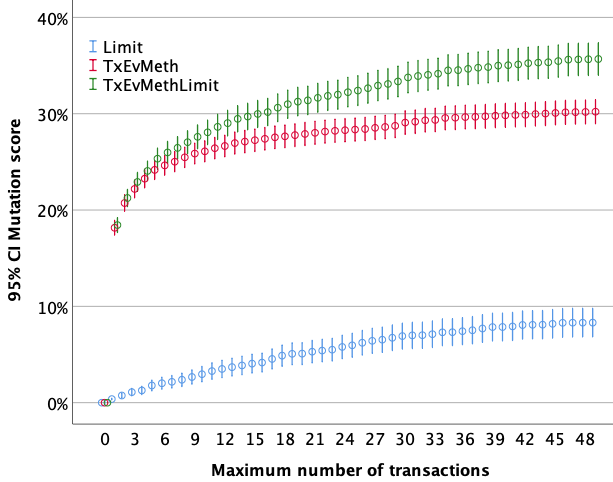}
      \caption{Percentage of non-equivalent mutants killed as a function of the length of the test. The error bars correspond to a confidence interval of 95\%.}
    \label{fig:mutation_score}
\end{figure}

{\bf Mutation score}
\autoref{fig:mutation_score} shows how the mutation score increases with the test size.
The error bars for a 95\% confidence interval are small.
The standard mutant killing condition {\it TxEvMeth} has most success early on, whereas the success of the {\it Limit} condition increases more gradually.
This difference can be explained as follows.
All tests execute the constructor method in transaction 0 and one regular method in transaction 1.
A large fraction of the tests only execute these two methods, hence most of the opportunity for killing a mutant on regular outputs occurs during transactions 0 and 1.
%

Since in general size matters~\cite{Inozemtseva2014}, we fixed the size of the tests to 50 transactions.
However, we could not fix the size of the smart contracts.
To study the influence of contract size we have calculated the rank correlation of the size of the bytecode and the mutation score.
For the combined killing condition {\it TxEvMethLimit}, we found Kendall's $\tau=-0.11$ ($p=0.01$, 2-tailed).
This means that the mutation score is not correlated with the size of the test.
We also calculated the correlation between the mutation score and the fraction of bytecodes that was executed by the test and found the correlation to be moderate: Kendall's $\tau=0.45$ ($p=0.01$, 2-tailed).
The literature reports similar figures~\cite{Gopinath2014}.

%

{\bf Efficiency of the mutation operators}
\autoref{tab:operator_stats} shows to what extent the mutation operators from \autoref{tab:operator_defs} have been successful.
The first column gives the name of the mutation operator.
The next two columns indicate how many mutants were not killed, and how many were killed.
The fourth column gives the total number of trivially non-equivalent mutants.
The last two columns give the percentages related to the numbers in columns two and three.
\begin{center}
\begin{table}
\caption{Contingency table of the mutation operators against the mutation score with the {\it TxEvMethLimit} mutant killing condition.}
\centering
\begin{tabular*}{0.61\textwidth}{l *{3}{r} | *{2}{r} }
\hline
\hline
Mutation	& \multicolumn{3}{c}{Non-equivalent mutants}	& \multicolumn{2}{c}{Percentage}\\
operator	& Not killed	& Killed	& Total		& Not killed	& Killed \\
\hline
AOR		&  2178		&  1039		&  3217		& 67.7\%	& 32.3\% \\
BOR		&  3549		&  2245		&  5794		& 61.3\%	& 38.7\% \\
ESD		&  5246		&  2595		&  7841		& 66.9\%	& 33.1\% \\
ITSCR		&  7796		&  4085		& 11881		& 65.6\%	& 34.4\% \\
JSRD		&  1866		&  1090		&  2956		& 63.1\%	& 36.9\% \\
LR\_A*		&    46		&    58		&   104		& 44.2\%	& 55.8\% \\
LR\_B		&   973		&   225		&  1198		& 81.2\%	& 18.8\% \\
LR\_I		&  1526		&  1089		&  2615		& 58.4\%	& 41.6\% \\
LR\_S		&   158		&   329		&   487		& 32.4\%	& 67.6\% \\
MORD*		&   921		&   143		&  1064		& 86.6\%	& 13.4\% \\
QRD*		&  1041		&  1761		&  2802		& 37.2\%	& 62.8\% \\
RAR*		&  3002		&  1158		&  4160		& 72.2\%	& 27.8\% \\
UORD		&   297		&   189		&   486		& 61.1\%	& 38.9\% \\
VDTSCS		&  2215		&  1050		&  3265		& 67.8\%	& 32.2\% \\
\hline
Total		& 30814		& 17056		& 47870		& 64.4\%	& 35.6\% \\
\hline
\hline
\multicolumn{6}{l}{$\chi^2=1759.6, df=13, p<0.001$}
\end{tabular*}
\label{tab:operator_stats}
\end{table}
\end{center}
\begin{center}
\begin{table}[ht]
\caption{Comparison of the effectiveness of all mutants versus the manually analysed mutant after test bootstrapping.}
\centering
\begin{tabular*}{0.715\textwidth}{l *{3}{r} | *{5}{r} }
\hline
\hline
Mutation       & \multicolumn{2}{c}{All mutants}	&& \multicolumn{4}{c}{Stratified sample} \\
operator       & Killed        & Total			&& Killed        & Total& Equivalent	& Killable \\
\hline
Mothra         & 35.1\%        & 39740			&& 41.3\%        & 223	& 3.1\%		& 96.9\% \\
Solidity       & 38.4\%        &  8130			&& 48.1\%        &  27	& 3.7\%		& 92.6\% \\
\hline
Total          & 35.6\%        & 47870			&& 42.0\%        & 250	& 3.2\%		& 96.4\% \\
\hline
\hline
\multicolumn{3}{l}{$\chi^2=32.2, df=1, p<0.001$}	&& \multicolumn{4}{l}{$\chi^2=8.3, df=2, p=0.016$}
\end{tabular*}
\label{tab:representative}
\end{table}
\end{center}
\begin{center}
\begin{table}[ht]
\caption{Comparison of coverage and mutation scores with related work.}
\centering
\begin{tabular*}{1\textwidth}{l @{\,} l l @{\,} l}
\hline
\hline
related			& average	& average	& DApps or smart contracts \\
work			& mutation	& statement	& \\
			& score		& coverage 	& \\
\hline
\cite{Honig2019} 	& 96.0\%	& 99.5\%	& Aragon OS,Openzeppelin-Solidity \\
\cite{Wu2019a}		& 43.9\%	& 68.9\%	& Skincoin,SmartIdentity,AirSwap,Cryptofin \\
\cite{Chapman2019}	& 40.3\%	& 95.4\%	& MetaCoin,MultiSigWallet,Alice \\
this			& 35.5\%	& 47.6\%*	& 963 Verified smart contracts \\
bootstr.		& 96.4\%	& 		& DBToken,MultiSigWallet,NumberBoard,casinoProxy,mall \\
\hline
\hline
\multicolumn{4}{l}{*) Average bytecode coverage}
\end{tabular*}
\label{tab:comparison}
\end{table}
\end{center}
\autoref{tab:operator_stats} shows that of the four Solidity specific operators (marked with an asterisk) {\it QRD*} is the most efficient when it comes to being easily killed. 
This is because subtle changes to the qualifiers, such as removing the \verb=payable= attribute from a method completely breaks the contract.
The standard {\it LR\_S} operator is the most efficient operator overall, because strings in Solidity are typically used for communication with the DApp built on top of a smart contract.
This means that even the smallest change to a string will be detected by comparing event parameters or method results.
The {\it MORD*} operator has the lowest efficiency (13.4\%), because relatively few historic transactions try to violate the access control implemented by the modifiers.
The {\it RAR*} operator also has a low efficiency, for the same reason: few historic transactions try to exploit bugs in address checking.

To assess the effect of using a replay test suite on the mutation score, we have analysed by hand all 250 mutants generated for 5 carefully selected smart contracts.
The analysis meant that for each mutant we looked at whether the test could be extended in such a way that the output could kill the test.
We call this {\it test bootstrapping}: a method that uses the replay tests to systematically create proper tests.
Each contract took us about one day to analyse, hence we had to limit the number of contracts to a small number like 5.

We used a stratified sampling method, taking one contract for the top 5\% contracts by mutation score, one contract from the bottom 5\%, and one contract each from $25\pm 2.5\%$, $50\pm2.5\%$ and $75\pm2.5\%$.
In selecting the contracts from the five ranges, we tried to avoid analysing the same type of contract more than once.
We analysed:
DBToken\footnote{DBToken on Etherscan \url{https://etherscan.io/address/0x42a952Ac23d020610355Cf425d0dfa58295287BE}},
MultiSigWallet\footnote{MultiSigWallet on Etherscan \url{https://etherscan.io/address/0xa723606e907bf84215d5785ea7f6cd93a0fbd121}},
the auction NumberBoard\footnote{NumberBoard on Etherscan \url{https://etherscan.io/address/0x9249133819102b2ed31680468c8c67F6Fe9E7505}},
the game casinoProxy\footnote{casinoProxy on Etherscan \url{https://etherscan.io/address/0x23a3db04432123ccdf6ef4459684329cc7c0b022}},
and the asset manager mall\footnote{mall on Etherscan \url{https://etherscan.io/address/0x3304a44aa16ec40fb53a5b8f086f230c237f683d}}.

\autoref{tab:representative} shows that the key statistics of the 250 manually analysed mutants and all 47,870 mutants are comparable, which suggests that the results we obtain for the 250 mutants are representative for all mutants.
The left and right half of \autoref{tab:representative} show a contingency table for the mutation operator types versus the status of the different mutation results.
The Mothra operators generate more semantically equivalent mutants than the solidity operators, but overall the percentage of semantically equivalent mutants in the stratified sample is only 3.2\%.
This indicates that semantically equivalent mutants do not inflate the base line statistics by more than a few per cent points.
The second conclusion that can be drawn from \autoref{tab:representative} is that almost 100\% mutation scores are possible with bootstrapped versions the replay tests.
A mutation score of 100\% is not achievable with the current implementation because it cannot detect the success or failure of an internal transaction.

Of the 3,304,002 bytecode instructions of all original contracts together, the replay tests execute 1,574,239 instructions, giving an average bytecode coverage of 47.6\%.
This makes achieving a high mutation score more difficult, as a mutation to code that is not executed will never be killed.
Our results are thus a baseline, and to explore how far removed the base line is from results with hand-crafted tests we compare our results to related work in \autoref{tab:comparison}.
The first column lists the citation to all related work that we aware of that report mutation scores for smart contracts.
The next two columns give the average mutation score, and the statement (bytecode) coverage.
The last column lists the DApps tested.

High statement coverage does not necessarily lead to a high mutation score, because the effect of the mutant may not be visible in the outputs.
Honig et al.~\cite{Honig2019} propose a small set of highly efficient mutants.
Chapman~\cite{Chapman2019} has about the same code coverage as Honig et al. but he proposes a large set of mutants; this reduces his mutation score.
Our main result is comparable to that of Wu et al.~\cite{Wu2019a}, but we have also shown that by taking the replay tests as a basis and improving them by test bootstrapping leads to the same high mutation scores that others have found.

{\bf Using outputs in a mutant killing condition}
The mutation score for {\it TxEvMethLimit} reaches 35.6\%, for {\it TxEvMeth} 30.2\%, and for {\it Limit} 8.3\%.
These mutation scores are low compared to state-of-the-art approaches~\cite{Papadakis2019}.
However, this figure is useful as a baseline for other approaches that use realistic tests instead of replay tests, and our aim was not to achieve a high mutation score, but to evaluate the applicability of mutation operators for a high number of contracts.
Moreover, even such a small score is helpful to show what kind of tests are missing and what has to be done to improve the tests.

{\bf Using gas in a mutant killing condition}
The contribution of the gas limit as killing condition is 35.6\%$-$30.2\%= 5.4\%, which seems rather small, but since it can require a huge manual effort to analyse the surviving mutant, even such a small fraction can save hours or days of manual work.

%% file: sec/discussion.tex
We put the results in a broader context and answer the research question.

{\bf Discarding stillborn mutants}
We believe that the percentage of the generated mutants that do not compile was relatively small and that it was acceptable in the exploratory context.
A production tool would have to implement more sophisticated mutations and would therefore have more knowledge of the semantics of smart contracts.

{\bf Discarding duplicate and equivalent mutants}
A simple but state-of-the-art approach has been used to address the equivalent mutant problem.
What we have not investigated is to what extent gas usage can be leveraged to discard more equivalent mutants; we suggest this as a topic for future work.

{\bf Mutation score}
By leveraging the historic data available on the Ethereum blockchain we have been able to generate tests that can be truncated to explore the relationship between test strength and the mutation score. As expected, the efficiency of the mutation operators tails off quickly.

{\bf Efficiency of the mutation operators}
One of the Solidity specific mutation operators was found to be more effective in producing mutants that have a high chance of being killed than most of the standard mutation operators.
We take this as an indication that further research is needed to develop more sophisticated Solidity specific mutation operators.
Another, important aspect of mutation operators is to which extent they can introduce common or severe bugs.
To assess this quality, we inspected the ability of our operators to introduce known vulnerabilities that can have severe consequences.
The associated vulnerabilities as (specified by the SWC) are shown in \autoref{tab:operator_defs}. It should be noted that we are able to simulate most vulnerabilities, which are related to simple mistakes in the source code. 
Additionally, it can be seen that the specific operators are concerned with more vulnerabilities and that they are more severe compared to the standard operators.
For example, {\it MORD*} can trigger three different kinds of access control related vulnerabilities, which would not be possible with the standard operators. 
This allowed us to discover a vulnerability in one of the contracts, which we have reported to the owners by way of responsible disclosure:
the modifier \verb=onlyOwner= is missing on one of the methods.

Our mutation score is lower than the scores reported by related work but this is not due to the choice of mutation operators but caused by the use of replay tests.
Bootstrapped replay tests are comparable to hand crafted tests.


{\bf Using outputs in a mutant killing condition}
Comparing the observable outputs of a transaction of mutant and original is an efficient killing condition.

{\bf Using gas in a mutant killing condition}
We have shown that using the gas consumption as a killing condition can improve the mutation score and hence the effectiveness of the mutation approach in general.
The contribution of the gas limit as a killing condition is small because gas limits are usually not tight.

We are now able to answer the main research question:
Our Solidity specific mutation operators are more efficient than the standard operators, and they also are able to introduce more and more severe vulnerabilities.
\autoref{tab:representative} shows that the difference in efficiency is 3.3 percent point, which is modest, but also statistically significant at ($p<0.001$).


%% file: sec/limitations.tex
%
%
A threat to the validity of our evaluation might be that we only consider a replay test suite that is less powerful than other testing techniques, which might obtain a higher mutation score.
Although there are better testing techniques, the focus of this work was not to find them, but to build a mutation-based test quality assurance method that can also serve as a baseline for other testing techniques.

Another argument regarding the validity of our method might be that it is not wise to kill a mutant only based on a different gas usage since it could still be semantically equal.
However, we believe that a different gas usage is still a valid reason to kill a mutant, because it represents a change in the monetary cost of a transaction.
Moreover, there are other comparable cost factors, like energy~\cite{DBLP:conf/sigsoft/JabbarvandM17} or execution time~\cite{Nilsson2006} that have been used as killing condition in the past.

The replication package of this paper\endnotemark[1] presents the recently proposed checklist\cite{Papadakis2019} for research on mutation testing to analyse our work.

%% file: sec/conclusions.tex
From almost 200 mutation operators from related work, we have generalized a compact set of 14 operators and tested them on a large scale. Our Solidity specific operators were able to produce nearly all the mutations that were proposed in the related work, with only a few minor exceptions.

To achieve scale, we used replay tests that were automatically generated from the Ethereum blockchain.
To the best of our knowledge there is no related work that performs mutation testing for smart contracts at scale.

The average mutation scores that we achieved with our replay tests were not as good as the scores from the best handwritten tests, but also various studies with manual tests have comparable scores. It should be pointed out that the score can depend strongly on the choice of the mutation operators, and manual tests often undergo many iterations to improve the score.
We have also shown that the replay tests can be improved manually, such that a score close to 100\% can be reached.

Using our novel killing condition based on the gas limit allowed us to improve the mutation score by a maximum of 5.5\%. This does not sound like much, but it can save a lot of manual effort for the analysis of surviving mutants.

Four of the 14 operators have been specifically developed for Solidity and the others originate form the core of the Mothra set.
The Solidity-specific operators are on average more efficient than the standard Mothra operators.
We have shown that serious vulnerabilities can be detected with the help of specific operators; this shows that tailor-made mutation operators are useful.

It would be interesting to study errors made by Solidity developers at scale to validate the mutation operators.
Another area of future work would be to use the gas limit on transactions to detect equivalent mutants.